\documentclass[%
 reprint,
superscriptaddress,
 amsmath,amssymb,
 aps,
prx,
nolongbibliography
]{revtex4-2}
\usepackage{placeins}
\usepackage{CJKutf8}
\usepackage{lineno}
\usepackage[colorlinks=true,linkcolor=blue]{hyperref}
\usepackage[squaren]{SIunits}
\usepackage{color}
\usepackage{amsmath}
\usepackage{caption,subcaption}
\usepackage{setspace}
\usepackage{multirow}
\usepackage{threeparttable}
\usepackage{booktabs}
\usepackage{graphicx}% Include figure files
\usepackage{dcolumn}% Align table columns on decimal point
\usepackage{bm}% bold math
\usepackage[T1]{fontenc}

\definecolor{corapink}{rgb}{0.97, 0.51, 0.47}

\definecolor{huablue}{rgb}{0.0, 0.3, 0.6}
\newcommand{\si}[1]{{\color{blue}#1}}
\newcommand\mpes{\si{S1}}

\newcommand\tsfour{\si{S3}}
\newcommand\bla{\si{S4}}

\newcommand\qitaxas{\si{S6}}
\newcommand\fengjianju{\si{S7}}
\newcommand\orbitalview{\si{S8}}
\begin{document}

\preprint{APS/123-QED}

%\title{Mapping Transient Structures of Cyclo[18]Carbon by Computational X-Ray Spectra}
%\title{Mapping Bond Alternation Dynamics in Cyclo[18]Carbon via Computational Core-Level Spectroscopy}

\title{Core-Level Spectroscopy Decodes Bond-Alternation Dynamics of Cyclo[18]Carbon}

\author{Minrui Wei} %Department of Applied Physics,
 \affiliation{MIIT Key Laboratory of Semiconductor Microstructure and Quantum Sensing, School of Physics, Nanjing University of Science and Technology, 210094 Nanjing, China}

\author{Zeyu Liu}
%\email{liuzy@just.edu.cn}
\affiliation{School of Environmental and Chemical Engineering, Jiangsu University of Science and Technology, 212100 Zhenjiang, China}

\author{Sheng-Yu Wang}
 \affiliation{MIIT Key Laboratory of Semiconductor Microstructure and Quantum Sensing, School of Physics, Nanjing University of Science and Technology, 210094 Nanjing, China}
 
\author{Jun-Rong Zhang}
 \affiliation{MIIT Key Laboratory of Semiconductor Microstructure and Quantum Sensing, School of Physics, Nanjing University of Science and Technology, 210094 Nanjing, China}
 
\author{Lu Zhang}
 \affiliation{MIIT Key Laboratory of Semiconductor Microstructure and Quantum Sensing, School of Physics, Nanjing University of Science and Technology, 210094 Nanjing, China}
 
\author{Guoyan Ge}
 \affiliation{MIIT Key Laboratory of Semiconductor Microstructure and Quantum Sensing, School of Physics, Nanjing University of Science and Technology, 210094 Nanjing, China}

\author{Weijie Hua}
 \email{wjhua@njust.edu.cn}
 \affiliation{MIIT Key Laboratory of Semiconductor Microstructure and Quantum Sensing, School of Physics, Nanjing University of Science and Technology, 210094 Nanjing, China}

\date{\today}
\clearpage

%\begin{CJK*}{UTF8}{gbsn}

\begin{abstract}
The advent of X-ray free-electron lasers and high-harmonic generation has made time-resolved X-ray spectroscopy a powerful tool for probing local atomic environments, yet whether localized core excitations can report on global collective distortions remains open. Cyclo[18]carbon (C$_{18}$), with its polyynic ground state (D$_\text{9h}$) and cumulenic transition state (D$_\text{18h}$), provides an ideal model to address this long-standing issue in bond-length alternation (BLA) dynamics. Mapping two-dimensional potential energy surfaces by first-principles simulations, we find that core ionization symmetrizes the ground-state double-well potential along the BLA coordinate. Our calculated X-ray spectra reveal remarkable sensitivity to bond-length variations: C1s ionization potentials vary by up to 2.4~eV across the BLA coordinate (1.1--1.4~\AA), with a 0.9~eV variation for minima predicted by different functionals, while NEXAFS $\pi^*$ peaks shift by up to 4~eV across the same coordinate. These predicted signatures provide a quantitative spectroscopy--structure dictionary for decoding transient structures in future ultrafast X-ray experiments and monitoring bond-alternation dynamics in real time.
\end{abstract}

\maketitle

%%%%%%%%%%%%%%%% FIGURE 1 %%%%%%%%%%%%%%%%%%%
\begin{figure*}[!htb]
  \centering
  \centering\includegraphics[width=15cm]{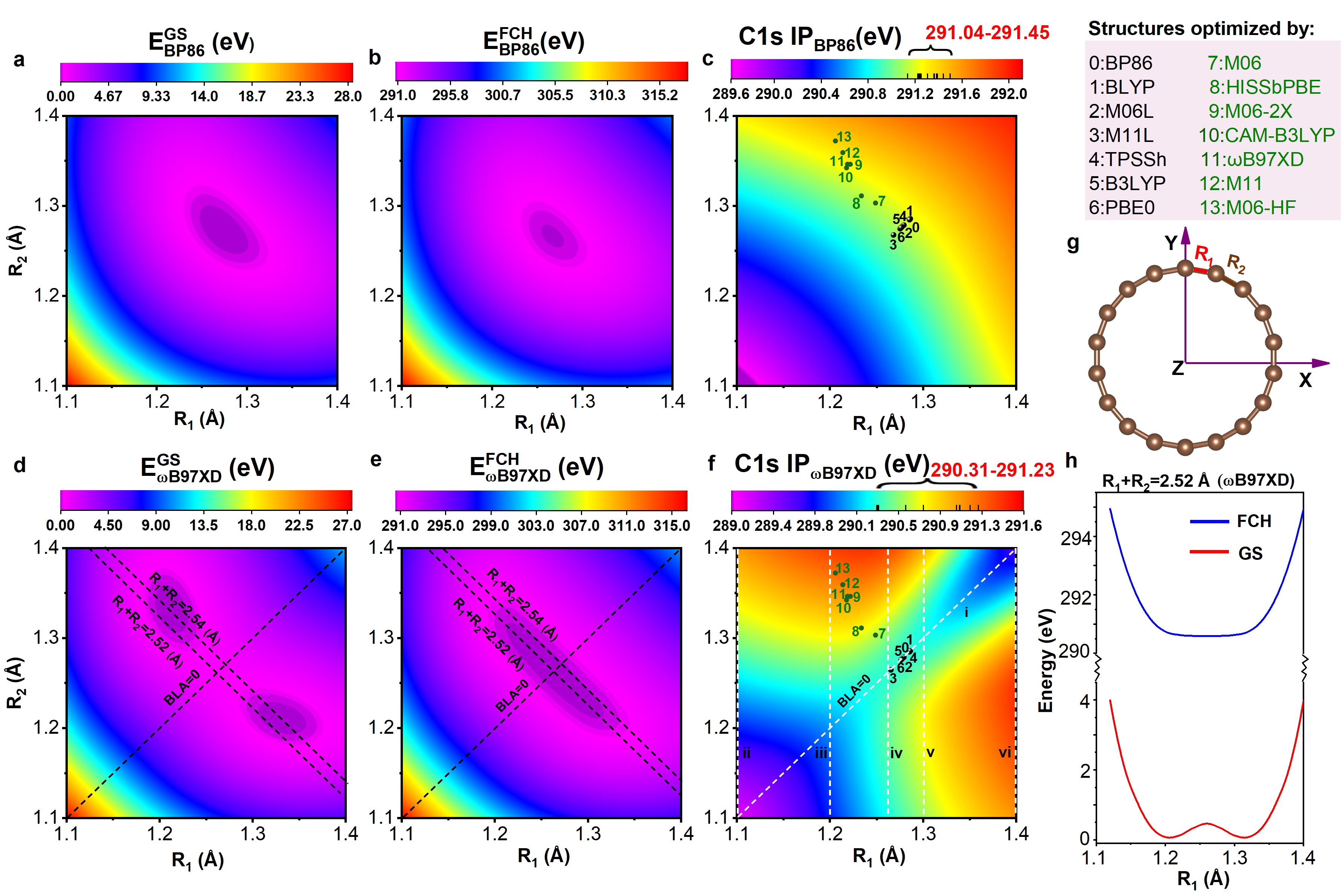}
  \caption{Core-ionization symmetrizes the potential energy landscape of C$_{18}$. (a-c) BP86 and (d-f) $\omega$B97XD results. (a,d) Ground-state energy $E^\text{GS}$, (b,e) C1s-ionized state energy $E^\text{FCH}$, and (c,f) vertical C1s ionization potential $I$ [see Supplementary Material, Eq. (\si{S1})]. Numbered points in (c,f) mark structures from Table \ref{tab:14func}, optimized with different functionals. (g) Definition of bond lengths $R_1$ and $R_2$. (h) PEC sections along $R_1+R_2=2.52$ {\AA} ($\omega$B97XD), showing the flattening of the potential upon core ionization.}
  \label{fig:ip}
\end{figure*}

Tracking molecular structural dynamics in real time is a central goal of ultrafast science. Time-resolved X-ray spectroscopies have emerged as powerful probes of local structural dynamics, owing to the localized nature of core excitations that interrogate specific atoms or bonds~\cite{bressler_molecular_2010, mukamel_multidimensional_2013, bergmann_using_2021, principi_preface_2024, wernet_orbital-specific_2015, jay_tracking_2023, ridente_femtosecond_2023, Gu2024, ostrom_probing_2015}.  Indeed, how local is ``local'' depends on the system: for structurally similar two-dimensional (2D) materials, the degree of nonlocality in core-level spectra correlates with the trend of $\pi$-electron delocalization (graphene $>$ C$_3$N $>$ h-BN)~\cite{zhang_choice_2022}. Beyond this localized picture, growing evidence suggests that X-ray spectra can encode information far beyond the immediate chemical environment. For example, core binding energies are found sensitive not only to the local bonding of the ionized atom but also to the broader electrostatic potential in which it resides \cite{Chambers2024}; localized fluorination of $\pi$-conjugated molecules\cite{Bischof2023} and nanosheets\cite{Xiao2024} can impact core binding energies far beyond the modification site, affecting the entire conjugated framework. These findings raise a broader question: can X-ray spectra be systematically decoded to extract global structural information? Cyclo[18]carbon (C$_{18}$) offers an ideal model system to address this question, as its 2D bond-length space and a bond-length alternation (BLA) coordinate describe a collective distortion that involves all 18 carbon atoms, providing a clear testbed for probing the interplay between local probes and global structural responses.

C$_{18}$, the first cyclocarbon synthesized on a surface in 2019~\cite{kaiser2019}, exhibits a fundamental structural dichotomy between polyynic (D$_\text{9h}$, alternating single and triple bonds) and cumulenic (D$_\text{18h}$, equal bonds) forms, with high-level theory firmly establishing the polyynic ground state~\cite{liusphybridizedbond2020, lu_accurate_2022} and bond-resolved atomic force microscopy confirming its alternating bond-length pattern~\cite{kaiser2019}. The cumulenic configuration of C$_{18}$ has been identified as a transition state (TS) connecting the two equivalent polyynic minima~\cite{baryshnikov_cyclo_2019, nandi_carbon_2020}, and the automerization between the two minima is predicted to be ultrafast, driven by carbon-atom quantum tunneling~\cite{nandi_carbon_2020}. Remarkably, even subtle variations in BLA induce significant shifts in optical absorption spectra~\cite{grillo_optical_2025}, underscoring the sensitivity of electronic excitations to the precise bond-length configuration. The cyclocarbon family has since expanded to include additional odd- and even-membered rings through on-surface synthesis~\cite{sun_onsurface_2023, gao_onsurface_2023, albrecht_odd-number_2024}, revealing how BLA varies with ring size and electronic structure. However, these prior investigations have largely focused on equilibrium structures and valence excitations; a systematic mapping of core-level spectral fingerprints across the full nuclear configuration space---essential for interpreting transient dynamics---has been lacking, and the transient cumulenic intermediate during BLA switching has never been directly observed (see recent reviews~\cite{cao_sp1-hybridized_2025, liu_cyclo18carbon_2025}). The importance of probing such structural dynamics has motivated theoretical strong-field studies of C$_{18}$~\cite{peng_solid_2023}. Capturing this fleeting state demands both ultrafast time resolution and a quantitative understanding of how core-level spectra encode the BLA coordinate.

The advent of X-ray free-electron lasers (XFELs) \cite{emma_first_2010} and high-harmonic generation (HHG) sources \cite{kapteyn_harnessing_2007} has enabled ultrafast X-ray spectra, such as femtosecond time-resolved XPS (TR-XPS)\cite{gabalski_time-resolved_2023} and time-resolved X-ray absorption spectroscopy (TR-XAS),\cite{ridente_femtosecond_2023, barlow_tracking_2024} which have been successfully applied to probe ultrafast processes. These developments make real-time tracking of BLA dynamics experimentally feasible. Yet a spectroscopy-structure dictionary that maps core-level features to specific ($R_1$, $R_2$) bond-length configurations remains absent. Establishing such a dictionary is challenging because the polyynic-cumulenic energy difference is small, requiring both accurate electronic structure methods and systematic sampling of the two-dimensional BLA coordinate. Moreover, core-hole creation introduces a strong perturbation that can dramatically alter the potential energy landscape, an effect not yet systematically explored for carbon rings. Without this theoretical foundation, future time-resolved measurements risk ambiguous interpretation.

To address these challenges, we performed comprehensive first-principles simulations by density functional theory (DFT) mainly with the range-separated hybrid functional $\omega$B97XD.~\cite{chai_long-range_2008} In the ground state, $\omega$B97XD was benchmarked against high-level wavefunction methods and shown to accurately describe C$_{18}$'s  BLA and electronic structure~\cite{lu_accurate_2022, liusphybridizedbond2020}. We constructed potential energy surfaces (PESs) for the ground, C1s-ionized, and C1s-excited states across the $R_1$--$R_2$ space (1.10--1.40~\AA). X-ray spectra were simulated by DFT with the full core hole (FCH)~\cite{triguero_separate_1999} and equivalent core hole (ECH)~\cite{jolly_thermodynamic_1970} approximations, which reliably reproduce K-edge spectra of carbon-based systems~\cite{zhang_accurate_2019, ge_qmmm_2022} and $\pi$-conjugated systems.~\cite{zhang_choice_2022, Xiao2024} Further computational details are provided in the Supplemental Material\cite{siC18}.

\begin{table}[!htbp]
\centering
\scalebox{0.85}{
\begin{threeparttable}
\caption{Calculated C1s ionization potentials ($I$, in eV) for C$_{18}$ using the FCH-DFT method with the BP86 and $\omega$B97XD functionals, evaluated at geometries optimized with various density functionals. Bond lengths $R_1$ and $R_2$ [defined in Fig.~\ref{fig:ip}(g)] and $\Delta R = R_2 - R_1$ are given in \AA.}\label{tab:14func}
\begin{tabular}{lllccccc}
\toprule
Index & Structure & $R_1$ & $R_2$ & $\Delta R$ &$I_\text{BP86}$ & $I_{\omega\text{B97XD}}$ \\
\midrule
D$_\text{18h}$ &&&&&&\\
0 & \textbf{min BP86} & 1.285 & 1.285 & 0.000 & 291.20 & 290.31 \\
1 & \textbf{min BLYP}  & 1.286 & 1.286 & 0.000 & 291.21 & 290.31 \\
2 & \textbf{min M06L} & 1.275 &1.275 & 0.000 & 291.12 & 290.33 \\
3 & \textbf{min M11L}  & 1.265 & 1.265 & 0.000 & 291.04 & 290.33 \\
4 & \textbf{min TPSSh}  & 1.279 & 1.279 & 0.000 & 291.15 & 290.32 \\
5 & \textbf{min B3LYP} & 1.277 & 1.277 & 0.000 & 291.13 & 290.32 \\
6 & \textbf{min PBE0}  & 1.276 & 1.276 & 0.000 & 291.13 & 290.32 \\
D$_\text{9h}$ &&&&&&\\
7 & \textbf{min M06} &  1.250 & 1.302 & 0.052 & 291.15 & 290.57 \\
8 & \textbf{min HISSbPBE}  & 1.235 & 1.310 & 0.075 & 291.14 & 290.73 \\
9 & \textbf{min M06-2X}& 1.223 & 1.345 & 0.122 & 291.29 & 291.06 \\
10 & \textbf{min CAM-B3LYP} & 1.219 & 1.341 & 0.122 & 291.26 & 291.03 \\
11 & \textbf{min $\boldsymbol{\omega}$B97XD} &  1.220 & 1.345 & 0.125 & 291.28 & 291.06 \\
12 & \textbf{min M11} & 1.215 & 1.358 & 0.143 & 291.34 & 291.15 \\
13 & \textbf{min M06-HF} & 1.207 & 1.371 & 0.164 & 291.40 & 291.23 \\
\bottomrule
\end{tabular}
\end{threeparttable}
}
\end{table}

%%%%%%%%%%%%%%%% FIGURE 2 %%%%%%%%%%%%%%%%%%%
\begin{figure*}[!htbp]
  \centering
  \includegraphics[width=12.5cm]{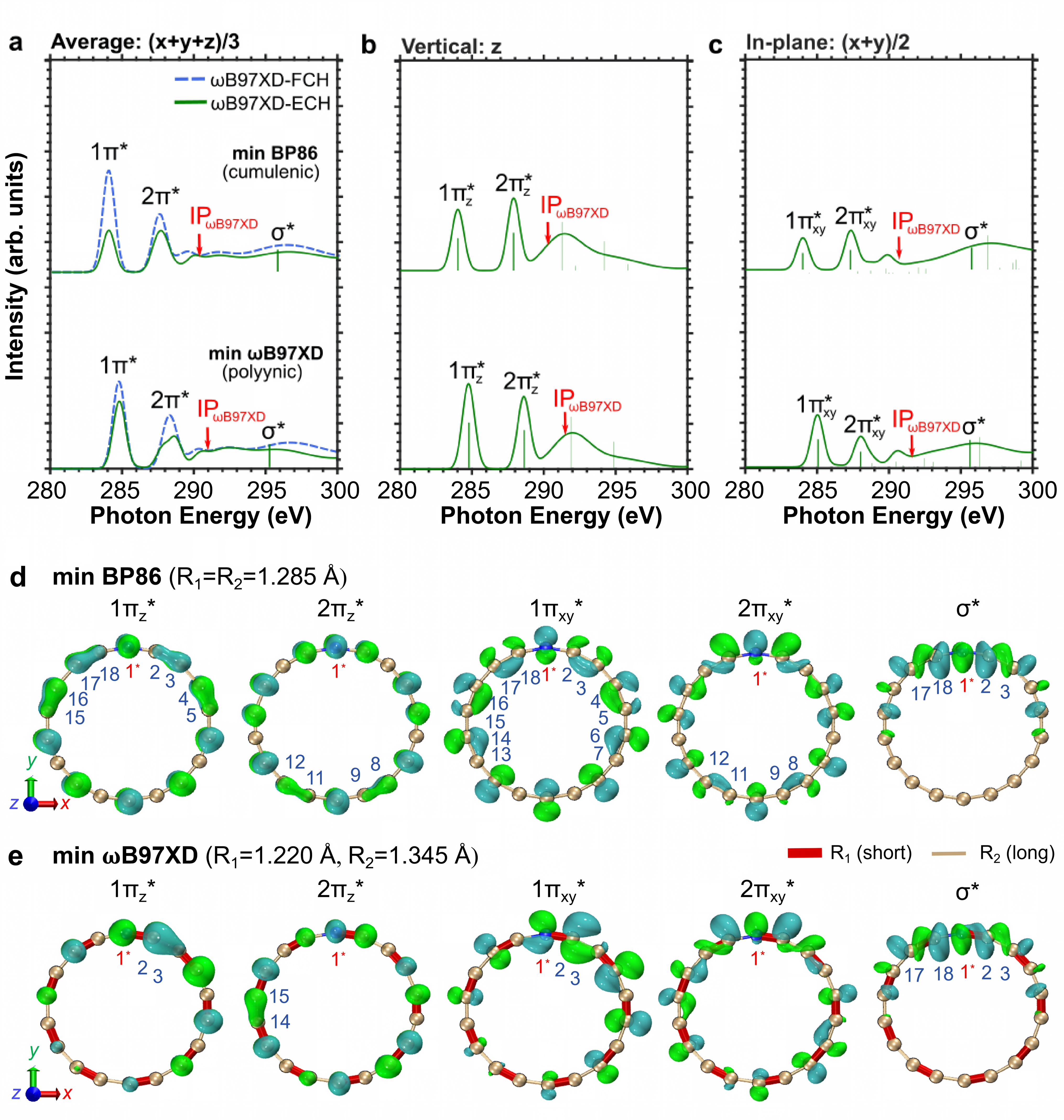}
\caption{Simulated $\omega$B97XD C1s NEXAFS spectra and final-state orbitals of C$_{18}$ for two representative geometries: the cumulenic (BP86-optimized) and polyynic ($\omega$B97XD-optimized) forms. (a) Orientationally-averaged spectra by ECH-DFT and FCH-DFT; (b) vertical and (c) in-plane components of the ECH spectra. (d-e) Corresponding final-state MOs (isovalue = 0.02) for major transitions in (b-c); see also Fig. {\orbitalview} for oblique views. Red arrows in (a-c) mark simulated IPs. In (d-e), C atoms are numbered clockwise, with red 1$^*$ indicating the core-excited site; navy numbers label bonding components, highlighting distinct orbital nodal patterns (see text for details).}
\label{fig:orbital}
\end{figure*}

Figure~\ref{fig:ip} presents the calculated 2D PESs of the ground and C1s-ionized states of C$_{18}$, and their energy difference, i.e., the vertical C1s ionization potential (IP). A direct comparison between the pure generalized gradient approximation (GGA) functional BP86 (a-c) and the range-separated hybrid $\omega$B97XD (d-f) reveals a critical methodological insight. For the ground state, BP86 erroneously predicts a single minimum at a cumulenic geometry ($R_1=R_2=1.28$~\AA) [Fig. \ref{fig:ip}(a)], whereas $\omega$B97XD correctly reproduces the polyynic ground state with two symmetry-equivalent minima at alternating bond lengths ($R_1,R_2$) = (1.22, 1.32) and (1.32, 1.22)~\AA\ [Fig. \ref{fig:ip}(d)], consistent with high-level wavefunction methods~\cite{lu_accurate_2022}. The $\omega$B97XD picture is robust: M06-2X yields similar PESs [Fig.~\si{\mpes}(a)], confirming this functional as appropriate for mapping C$_{18}$'s BLA coordinate.

The core-ionized state PES exhibits a dramatic reorganization. With $\omega$B97XD, the characteristic double well vanishes, replaced by an extended, oval-shaped region of low energy centered around $R_1=R_2$ [Fig. \ref{fig:ip}(e)]. This region has its long axis along $R_1+R_2=2.52$~\AA\ and its short axis along $R_1=R_2$, creating a broad, flat-bottomed potential where the two polyynic minima become indistinguishable [Fig. \ref{fig:ip}(h)]. In essence, core ionization symmetrizes the potential, stabilizing a continuum of cumulenic-like configurations rather than discrete minima. This symmetrization is particularly striking because it transforms the polyynic ground state into this extended symmetric region. In contrast, BP86 predicts single minima for both ground and ionized states [Figs. \ref{fig:ip}(a) and (b)], so no such transformation occurs. PEC slices along the long axis confirm the flat-bottom feature (Fig.~\si{\tsfour}), while those along the short axis show parabolic behavior with a minimum at (1.26, 1.26)~\AA\ (Fig.~\si{\bla}). The same topology is reproduced with M06-2X (Fig.~\si{\mpes}), confirming the symmetrization is not functional-specific. The symmetrization is reflected in the vertical IP maps [Figs. \ref{fig:ip}(c) and (f)]: while BP86 yields a monotonic IP increase, $\omega$B97XD predicts a complex, non-monotonic pattern with variations up to 2.35~eV, highlighting the exquisite sensitivity of the C1s binding energy to specific ($R_1$, $R_2$) pairs.

Figure \ref{fig:orbital}(a) presents the C1s NEXAFS spectra for two representative cumulenic and polyynic structures, simulated by $\omega$B97XD. The former with D$_\text{18h}$ symmetry ($R_1$=$R_2$=1.285 {\AA}) was obtained from BP86 optimization (labeled \textbf{min BP86}), and the latter with D$_\text{9h}$ symmetry [($R_1$, $R_2$)=(1.220, 1.345) {\AA}] from $\omega$B97XD optimization (\textbf{min $\boldsymbol{\omega}$B97XD}). For each structure, two complementary methods were employed: FCH provides more accurate spectral profiles, while ECH enables subsequent MO analysis~\cite{ge_qmmm_2022,zhang_accurate_2019}. Both spectra exhibit two $\pi^*$ resonances--an intense 1$\pi^*$ feature and a weaker 2$\pi^*$ peak--with a discernible $\sigma^*$ feature appearing at ca. 295~eV. The 1$\pi^*$ peak is blue-shifted by 0.7 eV from the cumulenic (284.05 eV) to the polyynic (284.73 eV) structure.

Beyond the magic-angle (isotropic) spectra, the planar geometry of C$_{18}$ allows access to polarization-dependent information. Figures \ref{fig:orbital}(b) and \ref{fig:orbital}(c) decompose the total spectra into out-of-plane ($z$) and in-plane ($x, y$) components, corresponding to incident X-ray polarization perpendicular and parallel to the molecular plane, respectively. Such decomposition is directly measurable in angle-resolved NEXAFS experiments on oriented samples.\cite{hua_x-ray_2010} Interestingly, for each $\pi^*$ resonance in the orientationally averaged spectra, the out-of-plane and in-plane components exhibit nearly identical energy positions and comparable intensities, indicating that the core-excited states retain a high degree of degeneracy despite the symmetry reduction upon core excitation.

Molecular orbital (MO) analysis reveals how the spatial distributions of the $\pi^*$ resonances encode structural information [Fig. \ref{fig:orbital}(d--e)]. For the cumulenic structure, which retains C$_\text{2v}$ symmetry in the core-excited state [Fig. \ref{fig:orbital}(d)], all five orbitals (four $\pi^*$ and one $\sigma^*$) are symmetric about the $yz$ plane. The four $\pi^*$ orbitals are delocalized over the entire molecule, whereas the $\sigma^*$ orbital is more localized near the core-excited site C1. In all these orbitals, two nodal planes separate C1 from its nearest neighbors C2 and C18. Each $\pi^*$ orbital exhibits a distinct nodal structure and can be described as a mixture of bonding and atomic character. For instance, $1\pi_z^*$ shows appreciable amplitude on the adjacent bonds (C2--C3, C4--C5, C15--C16, C17--C18), as well as on distant atoms C6, C8, C10, C12, and C14. In contrast, $2\pi_z^*$ displays a more atomic-like character, with density concentrated on selected sites (C2, C4, C5, C6, C8, C11, C14, C15, and C18) and bond amplitude surviving only on the remote bonds (C8--C9 and C11--C12). The two $\pi_{xy}^*$ orbitals exhibit nodal patterns generally similar to their $z$-plane counterparts.

In the polyynic structure [Fig. \ref{fig:orbital}(e)], the alternating bond lengths break the axial symmetry, reducing the core-excited state to the C$_\text{s}$ point group. Consequently, all MOs acquire distinct nodal patterns, although C1 consistently stands out with nodal planes between itself and its nearest neighbors C2 and C18. Compared with the cumulenic case, the four $\pi^*$ orbitals appear more localized, while the $\sigma^*$ orbital shows similar localization behavior. Bonding components are significantly reduced and are selectively concentrated on the long bonds. These systematic variations in orbital localization and nodal topology confirm that the shapes of the final-state $\pi^*$ orbitals serve as sensitive structural markers. The $\sigma^*$ orbitals of both structures exhibit stronger confinement near the core hole than their $\pi^*$ counterparts.

%%%%%%%%%%%%%%%% FIGURE 3 %%%%%%%%%%%%%%%%%%%
\begin{figure}[!htb]
  \centering
  \includegraphics[width=8.6cm]{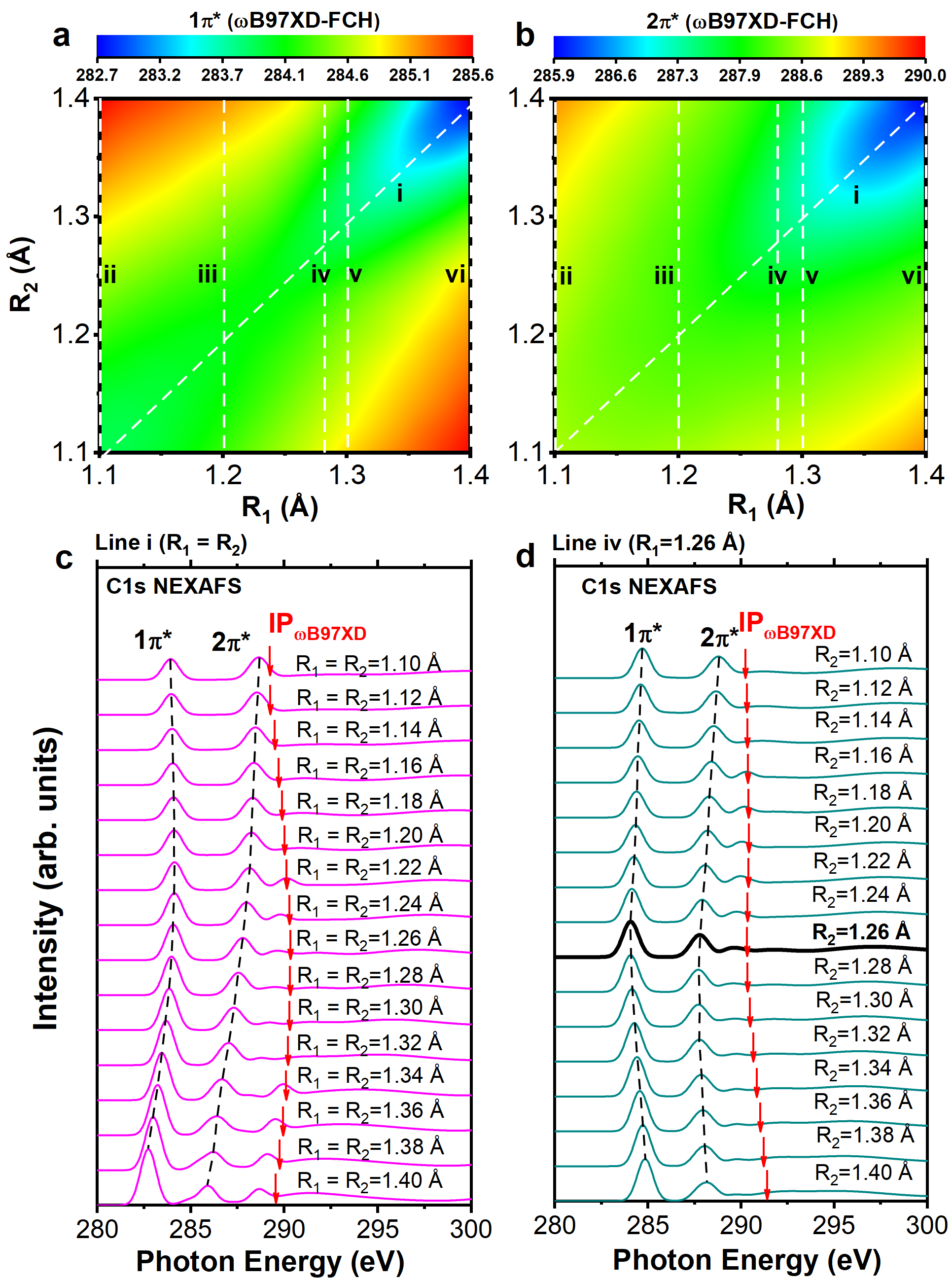}
  \caption{Simulated C1s NEXAFS spectra of C$_{18}$ by $\omega$B97XD-FCH. (a-b) Energy contour plots of the (a) 1${\pi}^*$ and (b) 2${\pi}^*$ peaks. (c-d) Orientationally averaged spectra along two representative paths: diagonal line i ($R_1=R_2$) and vertical line iv ($R_1=1.26$~\AA); see Fig. {\qitaxas} for lines ii-iii and v-vi. Red arrows mark simulated IPs.}
  \label{fig:xas}
\end{figure}

Figure \ref{fig:xas}(a)-(b) presents 2D maps of the energy positions of the 1$\pi^*$ and 2$\pi^*$ peaks as functions of bond lengths. Both resonances shift by $\sim$4~eV across the sampled ($R_1$, $R_2$) coordinate space (1.10--1.40~\AA), spanning 282--286~eV for 1$\pi^*$ and 286--290~eV for 2$\pi^*$, demonstrating a strong correlation between spectral features and structure. To examine this correlation, we slice the PESs along six representative paths: a diagonal line ($R_1 = R_2$, line i) and five vertical lines with $R_1$ fixed at 1.10, 1.20, 1.26, 1.30, and 1.40~\AA\ (lines ii--vi). Along the cumulenic line $R_1 = R_2$ [Fig.~\ref{fig:xas}(c)], increasing the bond length from 1.10 to 1.40~\AA\ monotonically redshifts 1$\pi^*$ and 2$\pi^*$ by $\sim$1.2~eV and $\sim$2.7~eV, respectively, with the redshift accelerating at larger $R$. In contrast, for fixed $R_1$ [Figs. \ref{fig:xas}(d) and \si{\qitaxas}], the dependence on $R_2$ is non-monotonic: the energy minima consistently occur at $R_1 = R_2$. This behavior exhibits two distinct regimes: when $R_1 > R_2$, increasing $R_2$ redshifts the peaks; when $R_1 < R_2$, further increasing $R_2$ induces a blueshift.

The peak separation $\delta = E(2\pi^*) - E(1\pi^*)$ provides an additional robust signature---independent of absolute energy calibration---that is particularly valuable for comparison with experimental spectra~\cite{ge_mapping_2024}. Along the $R_1 = R_2$ line (line i), $\delta$ decreases monotonically from 4.7 to 3.2~eV as the bond length increases from 1.10 to 1.40~\AA\ [Fig.~\si{\fengjianju}(a)], corresponding to a total reduction of 1.5~eV. Along fixed-$R_1$ lines (lines ii--vi), increasing $R_2$ from 1.10 to 1.40~\AA\ also systematically reduces $\delta$, with the reduction magnitude diminishing as $R_1$ increases: 0.85, 0.83, 0.80, 0.75, and 0.70~eV for $R_1 =$ 1.10, 1.20, 1.26, 1.30, and 1.40~\AA, respectively [Fig.~\si{\fengjianju}(b-f)].

In summary, we establish X-ray spectroscopy as a quantitative probe for the collective bond-alternation dynamics in C$_{18}$. Core ionization symmetrizes the ground-state double-well potential, stabilizing cumulenic configurations and reshaping the core-excited-state landscape. Both XPS and XAS exhibit distinct spectral fingerprints across the BLA coordinate: C1s ionization potentials vary by $\sim$0.9 eV for minima predicted by different functionals and up to 2.4~eV across the BLA coordinate (1.1--1.4~\AA), while $\pi^*$ peaks shift by up to $\sim$4 eV with characteristic non-monotonic energy shifts. These findings demonstrate that localized core excitations--though initiated at a single atomic site--carry global signatures of the entire carbon ring, embodying the interplay between local probes and extended structural responses. The resulting spectroscopy-structure dictionary provides a quantitative roadmap for future time-resolved experiments and is broadly applicable to other molecular systems with bond alternation or tautomerization.

%\section*{Author contributions}
%M.W. and Z.L. contributed equally to this work. W.H. conceived the idea and designed the research. M.W. performed the calculations. Z.L. provided expertise on the electronic structure methodology and background analysis of C$_{18}$. M.W. and W.H. wrote the manuscript with input from all authors. 

%\begin{acknowledgments}
\textit{Acknowledgments}--M.W. and Z.L. contributed equally to this work. We thank Professor Yi Luo for valuable discussions. This work was supported by the National Natural Science Foundation of China (Grant Nos. 12274229, 12504303), the National Postdoctoral Science Foundation of China (Grant No. 2025M774254), and the Excellent Postdoctoral Program of Jiangsu Province.
%\end{acknowledgments} 

\textit{Data Availability}--The data presented in this article is available online.\cite{wei_c18_data_2026}

\FloatBarrier

%\clearpage
%\bibliography{C18}

\begin{thebibliography}{38}%
\makeatletter
\providecommand \@ifxundefined [1]{%
 \@ifx{#1\undefined}
}%
\providecommand \@ifnum [1]{%
 \ifnum #1\expandafter \@firstoftwo
 \else \expandafter \@secondoftwo
 \fi
}%
\providecommand \@ifx [1]{%
 \ifx #1\expandafter \@firstoftwo
 \else \expandafter \@secondoftwo
 \fi
}%
\providecommand \natexlab [1]{#1}%
\providecommand \enquote  [1]{``#1''}%
\providecommand \bibnamefont  [1]{#1}%
\providecommand \bibfnamefont [1]{#1}%
\providecommand \citenamefont [1]{#1}%
\providecommand \href@noop [0]{\@secondoftwo}%
\providecommand \href [0]{\begingroup \@sanitize@url \@href}%
\providecommand \@href[1]{\@@startlink{#1}\@@href}%
\providecommand \@@href[1]{\endgroup#1\@@endlink}%
\providecommand \@sanitize@url [0]{\catcode `\\12\catcode `\$12\catcode `\&12\catcode `\#12\catcode `\^12\catcode `\_12\catcode `\%12\relax}%
\providecommand \@@startlink[1]{}%
\providecommand \@@endlink[0]{}%
\providecommand \url  [0]{\begingroup\@sanitize@url \@url }%
\providecommand \@url [1]{\endgroup\@href {#1}{\urlprefix }}%
\providecommand \urlprefix  [0]{URL }%
\providecommand \Eprint [0]{\href }%
\providecommand \doibase [0]{https://doi.org/}%
\providecommand \selectlanguage [0]{\@gobble}%
\providecommand \bibinfo  [0]{\@secondoftwo}%
\providecommand \bibfield  [0]{\@secondoftwo}%
\providecommand \translation [1]{[#1]}%
\providecommand \BibitemOpen [0]{}%
\providecommand \bibitemStop [0]{}%
\providecommand \bibitemNoStop [0]{.\EOS\space}%
\providecommand \EOS [0]{\spacefactor3000\relax}%
\providecommand \BibitemShut  [1]{\csname bibitem#1\endcsname}%
\let\auto@bib@innerbib\@empty
%</preamble>
\bibitem [{\citenamefont {Bressler}\ and\ \citenamefont {Chergui}(2010)}]{bressler_molecular_2010}%
  \BibitemOpen
  \bibfield  {author} {\bibinfo {author} {\bibfnamefont {C.}~\bibnamefont {Bressler}}\ and\ \bibinfo {author} {\bibfnamefont {M.}~\bibnamefont {Chergui}},\ }\href {https://doi.org/10.1146/annurev.physchem.012809.103353} {\bibfield  {journal} {\bibinfo  {journal} {Annu. Rev. Phys. Chem.}\ }\textbf {\bibinfo {volume} {61}},\ \bibinfo {pages} {263} (\bibinfo {year} {2010})}\BibitemShut {NoStop}%
\bibitem [{\citenamefont {Mukamel}\ \emph {et~al.}(2013)\citenamefont {Mukamel}, \citenamefont {Healion}, \citenamefont {Zhang},\ and\ \citenamefont {Biggs}}]{mukamel_multidimensional_2013}%
  \BibitemOpen
  \bibfield  {author} {\bibinfo {author} {\bibfnamefont {S.}~\bibnamefont {Mukamel}}, \bibinfo {author} {\bibfnamefont {D.}~\bibnamefont {Healion}}, \bibinfo {author} {\bibfnamefont {Y.}~\bibnamefont {Zhang}},\ and\ \bibinfo {author} {\bibfnamefont {J.~D.}\ \bibnamefont {Biggs}},\ }\href {https://doi.org/10.1146/annurev-physchem-040412-110021} {\bibfield  {journal} {\bibinfo  {journal} {Annu. Rev. Phys. Chem.}\ }\textbf {\bibinfo {volume} {64}},\ \bibinfo {pages} {101} (\bibinfo {year} {2013})}\BibitemShut {NoStop}%
\bibitem [{\citenamefont {Bergmann}\ \emph {et~al.}(2021)\citenamefont {Bergmann}, \citenamefont {Kern}, \citenamefont {Schoenlein}, \citenamefont {Wernet}, \citenamefont {Yachandra},\ and\ \citenamefont {Yano}}]{bergmann_using_2021}%
  \BibitemOpen
  \bibfield  {author} {\bibinfo {author} {\bibfnamefont {U.}~\bibnamefont {Bergmann}}, \bibinfo {author} {\bibfnamefont {J.}~\bibnamefont {Kern}}, \bibinfo {author} {\bibfnamefont {R.~W.}\ \bibnamefont {Schoenlein}}, \bibinfo {author} {\bibfnamefont {P.}~\bibnamefont {Wernet}}, \bibinfo {author} {\bibfnamefont {V.~K.}\ \bibnamefont {Yachandra}},\ and\ \bibinfo {author} {\bibfnamefont {J.}~\bibnamefont {Yano}},\ }\href {https://doi.org/10.1038/s42254-021-00289-3} {\bibfield  {journal} {\bibinfo  {journal} {Nat. Rev. Phys.}\ }\textbf {\bibinfo {volume} {3}},\ \bibinfo {pages} {264} (\bibinfo {year} {2021})}\BibitemShut {NoStop}%
\bibitem [{\citenamefont {Principi}(2024)}]{principi_preface_2024}%
  \BibitemOpen
  \bibfield  {author} {\bibinfo {author} {\bibfnamefont {E.}~\bibnamefont {Principi}},\ }\href {https://doi.org/10.1063/4.0000259} {\bibfield  {journal} {\bibinfo  {journal} {Struct. Dyn.}\ }\textbf {\bibinfo {volume} {11}},\ \bibinfo {pages} {030401} (\bibinfo {year} {2024})}\BibitemShut {NoStop}%
\bibitem [{\citenamefont {Wernet}\ \emph {et~al.}(2015)\citenamefont {Wernet}, \citenamefont {Kunnus}, \citenamefont {Josefsson}, \citenamefont {Rajkovic}, \citenamefont {Quevedo}, \citenamefont {Beye}, \citenamefont {Schreck}, \citenamefont {Gr{\"u}bel}, \citenamefont {Scholz}, \citenamefont {Nordlund}, \citenamefont {Zhang}, \citenamefont {Hartsock}, \citenamefont {Schlotter}, \citenamefont {Turner}, \citenamefont {Kennedy}, \citenamefont {Hennies}, \citenamefont {De~Groot}, \citenamefont {Gaffney}, \citenamefont {Techert}, \citenamefont {Odelius},\ and\ \citenamefont {F{\"o}hlisch}}]{wernet_orbital-specific_2015}%
  \BibitemOpen
  \bibfield  {author} {\bibinfo {author} {\bibfnamefont {P.}~\bibnamefont {Wernet}}, \bibinfo {author} {\bibfnamefont {K.}~\bibnamefont {Kunnus}}, \bibinfo {author} {\bibfnamefont {I.}~\bibnamefont {Josefsson}}, \bibinfo {author} {\bibfnamefont {I.}~\bibnamefont {Rajkovic}}, \bibinfo {author} {\bibfnamefont {W.}~\bibnamefont {Quevedo}}, \bibinfo {author} {\bibfnamefont {M.}~\bibnamefont {Beye}}, \bibinfo {author} {\bibfnamefont {S.}~\bibnamefont {Schreck}}, \bibinfo {author} {\bibfnamefont {S.}~\bibnamefont {Gr{\"u}bel}}, \bibinfo {author} {\bibfnamefont {M.}~\bibnamefont {Scholz}}, \bibinfo {author} {\bibfnamefont {D.}~\bibnamefont {Nordlund}}, \bibinfo {author} {\bibfnamefont {W.}~\bibnamefont {Zhang}}, \bibinfo {author} {\bibfnamefont {R.~W.}\ \bibnamefont {Hartsock}}, \bibinfo {author} {\bibfnamefont {W.~F.}\ \bibnamefont {Schlotter}}, \bibinfo {author} {\bibfnamefont {J.~J.}\ \bibnamefont {Turner}}, \bibinfo {author} {\bibfnamefont {B.}~\bibnamefont {Kennedy}}, \bibinfo {author} {\bibfnamefont
  {F.}~\bibnamefont {Hennies}}, \bibinfo {author} {\bibfnamefont {F.~M.~F.}\ \bibnamefont {De~Groot}}, \bibinfo {author} {\bibfnamefont {K.~J.}\ \bibnamefont {Gaffney}}, \bibinfo {author} {\bibfnamefont {S.}~\bibnamefont {Techert}}, \bibinfo {author} {\bibfnamefont {M.}~\bibnamefont {Odelius}},\ and\ \bibinfo {author} {\bibfnamefont {A.}~\bibnamefont {F{\"o}hlisch}},\ }\href {https://doi.org/10.1038/nature14296} {\bibfield  {journal} {\bibinfo  {journal} {Nature}\ }\textbf {\bibinfo {volume} {520}},\ \bibinfo {pages} {78} (\bibinfo {year} {2015})}\BibitemShut {NoStop}%
\bibitem [{\citenamefont {Jay}\ \emph {et~al.}(2023)\citenamefont {Jay}, \citenamefont {Banerjee}, \citenamefont {Leitner}, \citenamefont {Wang}, \citenamefont {Harich}, \citenamefont {Stefanuik}, \citenamefont {Wikmark}, \citenamefont {Coates}, \citenamefont {Beale}, \citenamefont {Kabanova}, \citenamefont {Kahraman}, \citenamefont {Wach}, \citenamefont {Ozerov}, \citenamefont {Arrell}, \citenamefont {Johnson}, \citenamefont {Borca}, \citenamefont {Cirelli}, \citenamefont {Bacellar}, \citenamefont {Milne}, \citenamefont {Huse}, \citenamefont {Smolentsev}, \citenamefont {Huthwelker}, \citenamefont {Odelius},\ and\ \citenamefont {Wernet}}]{jay_tracking_2023}%
  \BibitemOpen
  \bibfield  {author} {\bibinfo {author} {\bibfnamefont {R.~M.}\ \bibnamefont {Jay}}, \bibinfo {author} {\bibfnamefont {A.}~\bibnamefont {Banerjee}}, \bibinfo {author} {\bibfnamefont {T.}~\bibnamefont {Leitner}}, \bibinfo {author} {\bibfnamefont {R.-P.}\ \bibnamefont {Wang}}, \bibinfo {author} {\bibfnamefont {J.}~\bibnamefont {Harich}}, \bibinfo {author} {\bibfnamefont {R.}~\bibnamefont {Stefanuik}}, \bibinfo {author} {\bibfnamefont {H.}~\bibnamefont {Wikmark}}, \bibinfo {author} {\bibfnamefont {M.~R.}\ \bibnamefont {Coates}}, \bibinfo {author} {\bibfnamefont {E.~V.}\ \bibnamefont {Beale}}, \bibinfo {author} {\bibfnamefont {V.}~\bibnamefont {Kabanova}}, \bibinfo {author} {\bibfnamefont {A.}~\bibnamefont {Kahraman}}, \bibinfo {author} {\bibfnamefont {A.}~\bibnamefont {Wach}}, \bibinfo {author} {\bibfnamefont {D.}~\bibnamefont {Ozerov}}, \bibinfo {author} {\bibfnamefont {C.}~\bibnamefont {Arrell}}, \bibinfo {author} {\bibfnamefont {P.~J.~M.}\ \bibnamefont {Johnson}}, \bibinfo {author} {\bibfnamefont {C.~N.}\
  \bibnamefont {Borca}}, \bibinfo {author} {\bibfnamefont {C.}~\bibnamefont {Cirelli}}, \bibinfo {author} {\bibfnamefont {C.}~\bibnamefont {Bacellar}}, \bibinfo {author} {\bibfnamefont {C.}~\bibnamefont {Milne}}, \bibinfo {author} {\bibfnamefont {N.}~\bibnamefont {Huse}}, \bibinfo {author} {\bibfnamefont {G.}~\bibnamefont {Smolentsev}}, \bibinfo {author} {\bibfnamefont {T.}~\bibnamefont {Huthwelker}}, \bibinfo {author} {\bibfnamefont {M.}~\bibnamefont {Odelius}},\ and\ \bibinfo {author} {\bibfnamefont {P.}~\bibnamefont {Wernet}},\ }\href {https://doi.org/10.1126/science.adf8042} {\bibfield  {journal} {\bibinfo  {journal} {Science}\ }\textbf {\bibinfo {volume} {380}},\ \bibinfo {pages} {955} (\bibinfo {year} {2023})}\BibitemShut {NoStop}%
\bibitem [{\citenamefont {Ridente}\ \emph {et~al.}(2023)\citenamefont {Ridente}, \citenamefont {Hait}, \citenamefont {Haugen}, \citenamefont {Ross}, \citenamefont {Neumark}, \citenamefont {Head-Gordon},\ and\ \citenamefont {Leone}}]{ridente_femtosecond_2023}%
  \BibitemOpen
  \bibfield  {author} {\bibinfo {author} {\bibfnamefont {E.}~\bibnamefont {Ridente}}, \bibinfo {author} {\bibfnamefont {D.}~\bibnamefont {Hait}}, \bibinfo {author} {\bibfnamefont {E.~A.}\ \bibnamefont {Haugen}}, \bibinfo {author} {\bibfnamefont {A.~D.}\ \bibnamefont {Ross}}, \bibinfo {author} {\bibfnamefont {D.~M.}\ \bibnamefont {Neumark}}, \bibinfo {author} {\bibfnamefont {M.}~\bibnamefont {Head-Gordon}},\ and\ \bibinfo {author} {\bibfnamefont {S.~R.}\ \bibnamefont {Leone}},\ }\href {https://doi.org/10.1126/science.adg4421} {\bibfield  {journal} {\bibinfo  {journal} {Science}\ }\textbf {\bibinfo {volume} {380}},\ \bibinfo {pages} {713} (\bibinfo {year} {2023})}\BibitemShut {NoStop}%
\bibitem [{\citenamefont {Gu}\ \emph {et~al.}(2024)\citenamefont {Gu}, \citenamefont {Yong}, \citenamefont {Gu},\ and\ \citenamefont {Mukamel}}]{Gu2024}%
  \BibitemOpen
  \bibfield  {author} {\bibinfo {author} {\bibfnamefont {Y.}~\bibnamefont {Gu}}, \bibinfo {author} {\bibfnamefont {H.}~\bibnamefont {Yong}}, \bibinfo {author} {\bibfnamefont {B.}~\bibnamefont {Gu}},\ and\ \bibinfo {author} {\bibfnamefont {S.}~\bibnamefont {Mukamel}},\ }\href {https://doi.org/10.1073/pnas.2321343121} {\bibfield  {journal} {\bibinfo  {journal} {Proc. Natl. Acad. Sci. U.S.A.}\ }\textbf {\bibinfo {volume} {121}},\ \bibinfo {pages} {e2321343121} (\bibinfo {year} {2024})}\BibitemShut {NoStop}%
\bibitem [{\citenamefont {{\"O}str{\"o}m}\ \emph {et~al.}(2015)\citenamefont {{\"O}str{\"o}m}, \citenamefont {{\"O}berg}, \citenamefont {Xin}, \citenamefont {LaRue}, \citenamefont {Beye}, \citenamefont {Dell{\textquoteright}Angela}, \citenamefont {Gladh}, \citenamefont {Ng}, \citenamefont {Sellberg}, \citenamefont {Kaya}, \citenamefont {Mercurio}, \citenamefont {Nordlund}, \citenamefont {Hantschmann}, \citenamefont {Hieke}, \citenamefont {K{\"u}hn}, \citenamefont {Schlotter}, \citenamefont {Dakovski}, \citenamefont {Turner}, \citenamefont {Minitti}, \citenamefont {Mitra}, \citenamefont {Moeller}, \citenamefont {F{\"o}hlisch}, \citenamefont {Wolf}, \citenamefont {Wurth}, \citenamefont {Persson}, \citenamefont {N{\o}rskov}, \citenamefont {Abild-Pedersen}, \citenamefont {Ogasawara}, \citenamefont {Pettersson},\ and\ \citenamefont {Nilsson}}]{ostrom_probing_2015}%
  \BibitemOpen
  \bibfield  {author} {\bibinfo {author} {\bibfnamefont {H.}~\bibnamefont {{\"O}str{\"o}m}}, \bibinfo {author} {\bibfnamefont {H.}~\bibnamefont {{\"O}berg}}, \bibinfo {author} {\bibfnamefont {H.}~\bibnamefont {Xin}}, \bibinfo {author} {\bibfnamefont {J.}~\bibnamefont {LaRue}}, \bibinfo {author} {\bibfnamefont {M.}~\bibnamefont {Beye}}, \bibinfo {author} {\bibfnamefont {M.}~\bibnamefont {Dell{\textquoteright}Angela}}, \bibinfo {author} {\bibfnamefont {J.}~\bibnamefont {Gladh}}, \bibinfo {author} {\bibfnamefont {M.~L.}\ \bibnamefont {Ng}}, \bibinfo {author} {\bibfnamefont {J.~A.}\ \bibnamefont {Sellberg}}, \bibinfo {author} {\bibfnamefont {S.}~\bibnamefont {Kaya}}, \bibinfo {author} {\bibfnamefont {G.}~\bibnamefont {Mercurio}}, \bibinfo {author} {\bibfnamefont {D.}~\bibnamefont {Nordlund}}, \bibinfo {author} {\bibfnamefont {M.}~\bibnamefont {Hantschmann}}, \bibinfo {author} {\bibfnamefont {F.}~\bibnamefont {Hieke}}, \bibinfo {author} {\bibfnamefont {D.}~\bibnamefont {K{\"u}hn}}, \bibinfo {author} {\bibfnamefont
  {W.~F.}\ \bibnamefont {Schlotter}}, \bibinfo {author} {\bibfnamefont {G.~L.}\ \bibnamefont {Dakovski}}, \bibinfo {author} {\bibfnamefont {J.~J.}\ \bibnamefont {Turner}}, \bibinfo {author} {\bibfnamefont {M.~P.}\ \bibnamefont {Minitti}}, \bibinfo {author} {\bibfnamefont {A.}~\bibnamefont {Mitra}}, \bibinfo {author} {\bibfnamefont {S.~P.}\ \bibnamefont {Moeller}}, \bibinfo {author} {\bibfnamefont {A.}~\bibnamefont {F{\"o}hlisch}}, \bibinfo {author} {\bibfnamefont {M.}~\bibnamefont {Wolf}}, \bibinfo {author} {\bibfnamefont {W.}~\bibnamefont {Wurth}}, \bibinfo {author} {\bibfnamefont {M.}~\bibnamefont {Persson}}, \bibinfo {author} {\bibfnamefont {J.~K.}\ \bibnamefont {N{\o}rskov}}, \bibinfo {author} {\bibfnamefont {F.}~\bibnamefont {Abild-Pedersen}}, \bibinfo {author} {\bibfnamefont {H.}~\bibnamefont {Ogasawara}}, \bibinfo {author} {\bibfnamefont {L.~G.~M.}\ \bibnamefont {Pettersson}},\ and\ \bibinfo {author} {\bibfnamefont {A.}~\bibnamefont {Nilsson}},\ }\href {https://doi.org/10.1126/science.1261747}
  {\bibfield  {journal} {\bibinfo  {journal} {Science}\ }\textbf {\bibinfo {volume} {347}},\ \bibinfo {pages} {978} (\bibinfo {year} {2015})}\BibitemShut {NoStop}%
\bibitem [{\citenamefont {Zhang}\ \emph {et~al.}(2022)\citenamefont {Zhang}, \citenamefont {Wang}, \citenamefont {Ge}, \citenamefont {Wei}, \citenamefont {Hua},\ and\ \citenamefont {Ma}}]{zhang_choice_2022}%
  \BibitemOpen
  \bibfield  {author} {\bibinfo {author} {\bibfnamefont {J.-R.}\ \bibnamefont {Zhang}}, \bibinfo {author} {\bibfnamefont {S.-Y.}\ \bibnamefont {Wang}}, \bibinfo {author} {\bibfnamefont {G.}~\bibnamefont {Ge}}, \bibinfo {author} {\bibfnamefont {M.}~\bibnamefont {Wei}}, \bibinfo {author} {\bibfnamefont {W.}~\bibnamefont {Hua}},\ and\ \bibinfo {author} {\bibfnamefont {Y.}~\bibnamefont {Ma}},\ }\href@noop {} {\bibfield  {journal} {\bibinfo  {journal} {J. Chem. Phys.}\ }\textbf {\bibinfo {volume} {157}},\ \bibinfo {pages} {094704} (\bibinfo {year} {2022})}\BibitemShut {NoStop}%
\bibitem [{\citenamefont {Chambers}(2024)}]{Chambers2024}%
  \BibitemOpen
  \bibfield  {author} {\bibinfo {author} {\bibfnamefont {S.~A.}\ \bibnamefont {Chambers}},\ }\href {https://doi.org/10.1016/j.surfrep.2024.100638} {\bibfield  {journal} {\bibinfo  {journal} {Surf. Sci. Rep.}\ }\textbf {\bibinfo {volume} {79}},\ \bibinfo {pages} {100638} (\bibinfo {year} {2024})}\BibitemShut {NoStop}%
\bibitem [{\citenamefont {Bischof}\ \emph {et~al.}(2023)\citenamefont {Bischof}, \citenamefont {Radiev}, \citenamefont {Tripp}, \citenamefont {Hofmann}, \citenamefont {Geiger}, \citenamefont {Bettinger}, \citenamefont {Koert},\ and\ \citenamefont {Witte}}]{Bischof2023}%
  \BibitemOpen
  \bibfield  {author} {\bibinfo {author} {\bibfnamefont {D.}~\bibnamefont {Bischof}}, \bibinfo {author} {\bibfnamefont {Y.}~\bibnamefont {Radiev}}, \bibinfo {author} {\bibfnamefont {M.~W.}\ \bibnamefont {Tripp}}, \bibinfo {author} {\bibfnamefont {P.~E.}\ \bibnamefont {Hofmann}}, \bibinfo {author} {\bibfnamefont {T.}~\bibnamefont {Geiger}}, \bibinfo {author} {\bibfnamefont {H.~F.}\ \bibnamefont {Bettinger}}, \bibinfo {author} {\bibfnamefont {U.}~\bibnamefont {Koert}},\ and\ \bibinfo {author} {\bibfnamefont {G.}~\bibnamefont {Witte}},\ }\href {https://doi.org/10.1021/acs.jpclett.3c00287} {\bibfield  {journal} {\bibinfo  {journal} {J. Phys. Chem. Lett.}\ }\textbf {\bibinfo {volume} {14}},\ \bibinfo {pages} {2551} (\bibinfo {year} {2023})}\BibitemShut {NoStop}%
\bibitem [{\citenamefont {Xiao}\ \emph {et~al.}(2024)\citenamefont {Xiao}, \citenamefont {Zhang}, \citenamefont {Wang},\ and\ \citenamefont {Hua}}]{Xiao2024}%
  \BibitemOpen
  \bibfield  {author} {\bibinfo {author} {\bibfnamefont {Y.}~\bibnamefont {Xiao}}, \bibinfo {author} {\bibfnamefont {J.-R.}\ \bibnamefont {Zhang}}, \bibinfo {author} {\bibfnamefont {S.-Y.}\ \bibnamefont {Wang}},\ and\ \bibinfo {author} {\bibfnamefont {W.}~\bibnamefont {Hua}},\ }\href {https://doi.org/10.1021/prechem.4c00007} {\bibfield  {journal} {\bibinfo  {journal} {Precis. Chem.}\ }\textbf {\bibinfo {volume} {2}},\ \bibinfo {pages} {239} (\bibinfo {year} {2024})}\BibitemShut {NoStop}%
\bibitem [{\citenamefont {Kaiser}\ \emph {et~al.}(2019)\citenamefont {Kaiser}, \citenamefont {Scriven}, \citenamefont {Schulz}, \citenamefont {Gawel}, \citenamefont {Gross},\ and\ \citenamefont {Anderson}}]{kaiser2019}%
  \BibitemOpen
  \bibfield  {author} {\bibinfo {author} {\bibfnamefont {K.}~\bibnamefont {Kaiser}}, \bibinfo {author} {\bibfnamefont {L.~M.}\ \bibnamefont {Scriven}}, \bibinfo {author} {\bibfnamefont {F.}~\bibnamefont {Schulz}}, \bibinfo {author} {\bibfnamefont {P.}~\bibnamefont {Gawel}}, \bibinfo {author} {\bibfnamefont {L.}~\bibnamefont {Gross}},\ and\ \bibinfo {author} {\bibfnamefont {H.~L.}\ \bibnamefont {Anderson}},\ }\href {https://doi.org/10.1126/science.aay1914} {\bibfield  {journal} {\bibinfo  {journal} {Science}\ }\textbf {\bibinfo {volume} {365}},\ \bibinfo {pages} {1299} (\bibinfo {year} {2019})}\BibitemShut {NoStop}%
\bibitem [{\citenamefont {Liu}\ \emph {et~al.}(2020)\citenamefont {Liu}, \citenamefont {Lu},\ and\ \citenamefont {Chen}}]{liusphybridizedbond2020}%
  \BibitemOpen
  \bibfield  {author} {\bibinfo {author} {\bibfnamefont {Z.}~\bibnamefont {Liu}}, \bibinfo {author} {\bibfnamefont {T.}~\bibnamefont {Lu}},\ and\ \bibinfo {author} {\bibfnamefont {Q.}~\bibnamefont {Chen}},\ }\href {https://doi.org/10.1016/j.carbon.2020.04.099} {\bibfield  {journal} {\bibinfo  {journal} {Carbon}\ }\textbf {\bibinfo {volume} {165}},\ \bibinfo {pages} {468} (\bibinfo {year} {2020})}\BibitemShut {NoStop}%
\bibitem [{\citenamefont {Lu}\ \emph {et~al.}(2022)\citenamefont {Lu}, \citenamefont {Liu},\ and\ \citenamefont {Chen}}]{lu_accurate_2022}%
  \BibitemOpen
  \bibfield  {author} {\bibinfo {author} {\bibfnamefont {T.}~\bibnamefont {Lu}}, \bibinfo {author} {\bibfnamefont {Z.}~\bibnamefont {Liu}},\ and\ \bibinfo {author} {\bibfnamefont {Q.}~\bibnamefont {Chen}},\ }\href {https://doi.org/10.1088/1674-1056/ac873a} {\bibfield  {journal} {\bibinfo  {journal} {Chin. Phys. B}\ }\textbf {\bibinfo {volume} {31}},\ \bibinfo {pages} {126101} (\bibinfo {year} {2022})}\BibitemShut {NoStop}%
\bibitem [{\citenamefont {Baryshnikov}\ \emph {et~al.}(2019)\citenamefont {Baryshnikov}, \citenamefont {Valiev}, \citenamefont {Kuklin}, \citenamefont {Sundholm},\ and\ \citenamefont {{\AA}gren}}]{baryshnikov_cyclo_2019}%
  \BibitemOpen
  \bibfield  {author} {\bibinfo {author} {\bibfnamefont {G.~V.}\ \bibnamefont {Baryshnikov}}, \bibinfo {author} {\bibfnamefont {R.~R.}\ \bibnamefont {Valiev}}, \bibinfo {author} {\bibfnamefont {A.~V.}\ \bibnamefont {Kuklin}}, \bibinfo {author} {\bibfnamefont {D.}~\bibnamefont {Sundholm}},\ and\ \bibinfo {author} {\bibfnamefont {H.}~\bibnamefont {{\AA}gren}},\ }\href {https://doi.org/10.1021/acs.jpclett.9b02815} {\bibfield  {journal} {\bibinfo  {journal} {J. Phys. Chem. Lett.}\ }\textbf {\bibinfo {volume} {10}},\ \bibinfo {pages} {6701} (\bibinfo {year} {2019})}\BibitemShut {NoStop}%
\bibitem [{\citenamefont {Nandi}\ \emph {et~al.}(2020)\citenamefont {Nandi}, \citenamefont {Solel},\ and\ \citenamefont {Kozuch}}]{nandi_carbon_2020}%
  \BibitemOpen
  \bibfield  {author} {\bibinfo {author} {\bibfnamefont {A.}~\bibnamefont {Nandi}}, \bibinfo {author} {\bibfnamefont {E.}~\bibnamefont {Solel}},\ and\ \bibinfo {author} {\bibfnamefont {S.}~\bibnamefont {Kozuch}},\ }\href {https://doi.org/10.1002/chem.201904929} {\bibfield  {journal} {\bibinfo  {journal} {Chem. Eur. J.}\ }\textbf {\bibinfo {volume} {26}},\ \bibinfo {pages} {625} (\bibinfo {year} {2020})}\BibitemShut {NoStop}%
\bibitem [{\citenamefont {Grillo}\ \emph {et~al.}(2025)\citenamefont {Grillo}, \citenamefont {Pulci},\ and\ \citenamefont {Giovannini}}]{grillo_optical_2025}%
  \BibitemOpen
  \bibfield  {author} {\bibinfo {author} {\bibfnamefont {S.}~\bibnamefont {Grillo}}, \bibinfo {author} {\bibfnamefont {O.}~\bibnamefont {Pulci}},\ and\ \bibinfo {author} {\bibfnamefont {T.}~\bibnamefont {Giovannini}},\ }\href {https://doi.org/10.1039/D5SC05519A} {\bibfield  {journal} {\bibinfo  {journal} {Chem. Sci.}\ }\textbf {\bibinfo {volume} {16}},\ \bibinfo {pages} {22465} (\bibinfo {year} {2025})}\BibitemShut {NoStop}%
\bibitem [{\citenamefont {Sun}\ \emph {et~al.}(2023)\citenamefont {Sun}, \citenamefont {Zheng}, \citenamefont {Gao}, \citenamefont {Kang}, \citenamefont {Zhao},\ and\ \citenamefont {Xu}}]{sun_onsurface_2023}%
  \BibitemOpen
  \bibfield  {author} {\bibinfo {author} {\bibfnamefont {L.}~\bibnamefont {Sun}}, \bibinfo {author} {\bibfnamefont {W.}~\bibnamefont {Zheng}}, \bibinfo {author} {\bibfnamefont {W.}~\bibnamefont {Gao}}, \bibinfo {author} {\bibfnamefont {F.}~\bibnamefont {Kang}}, \bibinfo {author} {\bibfnamefont {M.}~\bibnamefont {Zhao}},\ and\ \bibinfo {author} {\bibfnamefont {W.}~\bibnamefont {Xu}},\ }\href {https://doi.org/10.1038/s41586-023-06741-x} {\bibfield  {journal} {\bibinfo  {journal} {Nature}\ }\textbf {\bibinfo {volume} {623}},\ \bibinfo {pages} {972} (\bibinfo {year} {2023})}\BibitemShut {NoStop}%
\bibitem [{\citenamefont {Gao}\ \emph {et~al.}(2023)\citenamefont {Gao}, \citenamefont {Albrecht}, \citenamefont {Ron{\v c}evi{\'c}}, \citenamefont {Ettedgui}, \citenamefont {Kumar}, \citenamefont {Scriven}, \citenamefont {Christensen}, \citenamefont {Mishra}, \citenamefont {Righetti}, \citenamefont {Rossmannek}, \citenamefont {Tavernelli}, \citenamefont {Anderson},\ and\ \citenamefont {Gross}}]{gao_onsurface_2023}%
  \BibitemOpen
  \bibfield  {author} {\bibinfo {author} {\bibfnamefont {Y.}~\bibnamefont {Gao}}, \bibinfo {author} {\bibfnamefont {F.}~\bibnamefont {Albrecht}}, \bibinfo {author} {\bibfnamefont {I.}~\bibnamefont {Ron{\v c}evi{\'c}}}, \bibinfo {author} {\bibfnamefont {I.}~\bibnamefont {Ettedgui}}, \bibinfo {author} {\bibfnamefont {P.}~\bibnamefont {Kumar}}, \bibinfo {author} {\bibfnamefont {L.~M.}\ \bibnamefont {Scriven}}, \bibinfo {author} {\bibfnamefont {K.~E.}\ \bibnamefont {Christensen}}, \bibinfo {author} {\bibfnamefont {S.}~\bibnamefont {Mishra}}, \bibinfo {author} {\bibfnamefont {L.}~\bibnamefont {Righetti}}, \bibinfo {author} {\bibfnamefont {M.}~\bibnamefont {Rossmannek}}, \bibinfo {author} {\bibfnamefont {I.}~\bibnamefont {Tavernelli}}, \bibinfo {author} {\bibfnamefont {H.~L.}\ \bibnamefont {Anderson}},\ and\ \bibinfo {author} {\bibfnamefont {L.}~\bibnamefont {Gross}},\ }\href {https://www.nature.com/articles/s41586-023-06566-8} {\bibfield  {journal} {\bibinfo  {journal} {Nature}\ }\textbf {\bibinfo {volume}
  {623}},\ \bibinfo {pages} {977} (\bibinfo {year} {2023})}\BibitemShut {NoStop}%
\bibitem [{\citenamefont {Albrecht}\ \emph {et~al.}(2024)\citenamefont {Albrecht}, \citenamefont {Ron{\v c}evi{\'c}}, \citenamefont {Gao}, \citenamefont {Paschke}, \citenamefont {Baiardi}, \citenamefont {Tavernelli}, \citenamefont {Mishra}, \citenamefont {Anderson},\ and\ \citenamefont {Gross}}]{albrecht_odd-number_2024}%
  \BibitemOpen
  \bibfield  {author} {\bibinfo {author} {\bibfnamefont {F.}~\bibnamefont {Albrecht}}, \bibinfo {author} {\bibfnamefont {I.}~\bibnamefont {Ron{\v c}evi{\'c}}}, \bibinfo {author} {\bibfnamefont {Y.}~\bibnamefont {Gao}}, \bibinfo {author} {\bibfnamefont {F.}~\bibnamefont {Paschke}}, \bibinfo {author} {\bibfnamefont {A.}~\bibnamefont {Baiardi}}, \bibinfo {author} {\bibfnamefont {I.}~\bibnamefont {Tavernelli}}, \bibinfo {author} {\bibfnamefont {S.}~\bibnamefont {Mishra}}, \bibinfo {author} {\bibfnamefont {H.~L.}\ \bibnamefont {Anderson}},\ and\ \bibinfo {author} {\bibfnamefont {L.}~\bibnamefont {Gross}},\ }\href {https://doi.org/10.1126/science.ado1399} {\bibfield  {journal} {\bibinfo  {journal} {Science}\ }\textbf {\bibinfo {volume} {384}},\ \bibinfo {pages} {677} (\bibinfo {year} {2024})}\BibitemShut {NoStop}%
\bibitem [{\citenamefont {Cao}\ and\ \citenamefont {Shi}(2025)}]{cao_sp1-hybridized_2025}%
  \BibitemOpen
  \bibfield  {author} {\bibinfo {author} {\bibfnamefont {H.}~\bibnamefont {Cao}}\ and\ \bibinfo {author} {\bibfnamefont {L.}~\bibnamefont {Shi}},\ }\href {https://doi.org/10.1016/j.cclet.2024.110466} {\bibfield  {journal} {\bibinfo  {journal} {Chin. Chem. Lett.}\ }\textbf {\bibinfo {volume} {36}},\ \bibinfo {pages} {110466} (\bibinfo {year} {2025})}\BibitemShut {NoStop}%
\bibitem [{\citenamefont {Liu}\ and\ \citenamefont {Lu}(2025)}]{liu_cyclo18carbon_2025}%
  \BibitemOpen
  \bibfield  {author} {\bibinfo {author} {\bibfnamefont {Z.}~\bibnamefont {Liu}}\ and\ \bibinfo {author} {\bibfnamefont {T.}~\bibnamefont {Lu}},\ }\href {https://doi.org/10.1021/accountsmr.5c00131} {\bibfield  {journal} {\bibinfo  {journal} {Acc. Mater. Res.}\ }\textbf {\bibinfo {volume} {6}},\ \bibinfo {pages} {1220} (\bibinfo {year} {2025})}\BibitemShut {NoStop}%
\bibitem [{\citenamefont {Peng}\ \emph {et~al.}(2023)\citenamefont {Peng}, \citenamefont {Wu}, \citenamefont {Yuan}, \citenamefont {Chi}, \citenamefont {Jiang}, \citenamefont {Dorfman}, \citenamefont {Yu},\ and\ \citenamefont {Lu}}]{peng_solid_2023}%
  \BibitemOpen
  \bibfield  {author} {\bibinfo {author} {\bibfnamefont {Y.}~\bibnamefont {Peng}}, \bibinfo {author} {\bibfnamefont {T.}~\bibnamefont {Wu}}, \bibinfo {author} {\bibfnamefont {G.}~\bibnamefont {Yuan}}, \bibinfo {author} {\bibfnamefont {L.}~\bibnamefont {Chi}}, \bibinfo {author} {\bibfnamefont {S.}~\bibnamefont {Jiang}}, \bibinfo {author} {\bibfnamefont {K.}~\bibnamefont {Dorfman}}, \bibinfo {author} {\bibfnamefont {C.}~\bibnamefont {Yu}},\ and\ \bibinfo {author} {\bibfnamefont {R.}~\bibnamefont {Lu}},\ }\href {https://www.science.org/doi/10.1126/sciadv.add6810} {\bibfield  {journal} {\bibinfo  {journal} {Sci. Adv.}\ }\textbf {\bibinfo {volume} {9}},\ \bibinfo {pages} {eadd6810} (\bibinfo {year} {2023})}\BibitemShut {NoStop}%
\bibitem [{\citenamefont {Emma}\ \emph {et~al.}(2010)\citenamefont {Emma}, \citenamefont {Akre}, \citenamefont {Arthur}, \citenamefont {Bionta}, \citenamefont {Bostedt}, \citenamefont {Bozek}, \citenamefont {Brachmann}, \citenamefont {Bucksbaum}, \citenamefont {Coffee}, \citenamefont {Decker}, \citenamefont {Ding}, \citenamefont {Dowell}, \citenamefont {Edstrom}, \citenamefont {Fisher}, \citenamefont {Frisch}, \citenamefont {Gilevich}, \citenamefont {Hastings}, \citenamefont {Hays}, \citenamefont {Hering}, \citenamefont {Huang}, \citenamefont {Iverson}, \citenamefont {Loos}, \citenamefont {Messerschmidt}, \citenamefont {Miahnahri}, \citenamefont {Moeller}, \citenamefont {Nuhn}, \citenamefont {Pile}, \citenamefont {Ratner}, \citenamefont {Rzepiela}, \citenamefont {Schultz}, \citenamefont {Smith}, \citenamefont {Stefan}, \citenamefont {Tompkins}, \citenamefont {Turner}, \citenamefont {Welch}, \citenamefont {White}, \citenamefont {Wu}, \citenamefont {Yocky},\ and\ \citenamefont {Galayda}}]{emma_first_2010}%
  \BibitemOpen
  \bibfield  {author} {\bibinfo {author} {\bibfnamefont {P.}~\bibnamefont {Emma}}, \bibinfo {author} {\bibfnamefont {R.}~\bibnamefont {Akre}}, \bibinfo {author} {\bibfnamefont {J.}~\bibnamefont {Arthur}}, \bibinfo {author} {\bibfnamefont {R.}~\bibnamefont {Bionta}}, \bibinfo {author} {\bibfnamefont {C.}~\bibnamefont {Bostedt}}, \bibinfo {author} {\bibfnamefont {J.}~\bibnamefont {Bozek}}, \bibinfo {author} {\bibfnamefont {A.}~\bibnamefont {Brachmann}}, \bibinfo {author} {\bibfnamefont {P.}~\bibnamefont {Bucksbaum}}, \bibinfo {author} {\bibfnamefont {R.}~\bibnamefont {Coffee}}, \bibinfo {author} {\bibfnamefont {F.-J.}\ \bibnamefont {Decker}}, \bibinfo {author} {\bibfnamefont {Y.}~\bibnamefont {Ding}}, \bibinfo {author} {\bibfnamefont {D.}~\bibnamefont {Dowell}}, \bibinfo {author} {\bibfnamefont {S.}~\bibnamefont {Edstrom}}, \bibinfo {author} {\bibfnamefont {A.}~\bibnamefont {Fisher}}, \bibinfo {author} {\bibfnamefont {J.}~\bibnamefont {Frisch}}, \bibinfo {author} {\bibfnamefont {S.}~\bibnamefont {Gilevich}},
  \bibinfo {author} {\bibfnamefont {J.}~\bibnamefont {Hastings}}, \bibinfo {author} {\bibfnamefont {G.}~\bibnamefont {Hays}}, \bibinfo {author} {\bibfnamefont {P.}~\bibnamefont {Hering}}, \bibinfo {author} {\bibfnamefont {Z.}~\bibnamefont {Huang}}, \bibinfo {author} {\bibfnamefont {R.}~\bibnamefont {Iverson}}, \bibinfo {author} {\bibfnamefont {H.}~\bibnamefont {Loos}}, \bibinfo {author} {\bibfnamefont {M.}~\bibnamefont {Messerschmidt}}, \bibinfo {author} {\bibfnamefont {A.}~\bibnamefont {Miahnahri}}, \bibinfo {author} {\bibfnamefont {S.}~\bibnamefont {Moeller}}, \bibinfo {author} {\bibfnamefont {H.-D.}\ \bibnamefont {Nuhn}}, \bibinfo {author} {\bibfnamefont {G.}~\bibnamefont {Pile}}, \bibinfo {author} {\bibfnamefont {D.}~\bibnamefont {Ratner}}, \bibinfo {author} {\bibfnamefont {J.}~\bibnamefont {Rzepiela}}, \bibinfo {author} {\bibfnamefont {D.}~\bibnamefont {Schultz}}, \bibinfo {author} {\bibfnamefont {T.}~\bibnamefont {Smith}}, \bibinfo {author} {\bibfnamefont {P.}~\bibnamefont {Stefan}}, \bibinfo {author}
  {\bibfnamefont {H.}~\bibnamefont {Tompkins}}, \bibinfo {author} {\bibfnamefont {J.}~\bibnamefont {Turner}}, \bibinfo {author} {\bibfnamefont {J.}~\bibnamefont {Welch}}, \bibinfo {author} {\bibfnamefont {W.}~\bibnamefont {White}}, \bibinfo {author} {\bibfnamefont {J.}~\bibnamefont {Wu}}, \bibinfo {author} {\bibfnamefont {G.}~\bibnamefont {Yocky}},\ and\ \bibinfo {author} {\bibfnamefont {J.}~\bibnamefont {Galayda}},\ }\href {https://doi.org/10.1038/nphoton.2010.176} {\bibfield  {journal} {\bibinfo  {journal} {Nat. Photonics}\ }\textbf {\bibinfo {volume} {4}},\ \bibinfo {pages} {641} (\bibinfo {year} {2010})}\BibitemShut {NoStop}%
\bibitem [{\citenamefont {Kapteyn}\ \emph {et~al.}(2007)\citenamefont {Kapteyn}, \citenamefont {Cohen}, \citenamefont {Christov},\ and\ \citenamefont {Murnane}}]{kapteyn_harnessing_2007}%
  \BibitemOpen
  \bibfield  {author} {\bibinfo {author} {\bibfnamefont {H.}~\bibnamefont {Kapteyn}}, \bibinfo {author} {\bibfnamefont {O.}~\bibnamefont {Cohen}}, \bibinfo {author} {\bibfnamefont {I.}~\bibnamefont {Christov}},\ and\ \bibinfo {author} {\bibfnamefont {M.}~\bibnamefont {Murnane}},\ }\href {https://doi.org/10.1126/science.1143679} {\bibfield  {journal} {\bibinfo  {journal} {Science}\ }\textbf {\bibinfo {volume} {317}},\ \bibinfo {pages} {775} (\bibinfo {year} {2007})}\BibitemShut {NoStop}%
\bibitem [{\citenamefont {Gabalski}\ \emph {et~al.}(2023)\citenamefont {Gabalski}, \citenamefont {Allum}, \citenamefont {Seidu}, \citenamefont {Britton}, \citenamefont {Brenner}, \citenamefont {Bromberger}, \citenamefont {Brouard}, \citenamefont {Bucksbaum}, \citenamefont {Burt}, \citenamefont {Cryan}, \citenamefont {Driver}, \citenamefont {Ekanayake}, \citenamefont {Erk}, \citenamefont {Garg}, \citenamefont {Gougoula}, \citenamefont {Heathcote}, \citenamefont {Hockett}, \citenamefont {Holland}, \citenamefont {Howard}, \citenamefont {Kumar}, \citenamefont {Lee}, \citenamefont {Li}, \citenamefont {McManus}, \citenamefont {Mikosch}, \citenamefont {Milesevic}, \citenamefont {Minns}, \citenamefont {Neville}, \citenamefont {{Atia-Tul-Noor}}, \citenamefont {Papadopoulou}, \citenamefont {Passow}, \citenamefont {Razmus}, \citenamefont {R{\"o}der}, \citenamefont {Rouz{\'e}e}, \citenamefont {Simao}, \citenamefont {Unwin}, \citenamefont {Vallance}, \citenamefont {Walmsley}, \citenamefont {Wang}, \citenamefont {Rolles},
  \citenamefont {Stolow}, \citenamefont {Schuurman},\ and\ \citenamefont {Forbes}}]{gabalski_time-resolved_2023}%
  \BibitemOpen
  \bibfield  {author} {\bibinfo {author} {\bibfnamefont {I.}~\bibnamefont {Gabalski}}, \bibinfo {author} {\bibfnamefont {F.}~\bibnamefont {Allum}}, \bibinfo {author} {\bibfnamefont {I.}~\bibnamefont {Seidu}}, \bibinfo {author} {\bibfnamefont {M.}~\bibnamefont {Britton}}, \bibinfo {author} {\bibfnamefont {G.}~\bibnamefont {Brenner}}, \bibinfo {author} {\bibfnamefont {H.}~\bibnamefont {Bromberger}}, \bibinfo {author} {\bibfnamefont {M.}~\bibnamefont {Brouard}}, \bibinfo {author} {\bibfnamefont {P.~H.}\ \bibnamefont {Bucksbaum}}, \bibinfo {author} {\bibfnamefont {M.}~\bibnamefont {Burt}}, \bibinfo {author} {\bibfnamefont {J.~P.}\ \bibnamefont {Cryan}}, \bibinfo {author} {\bibfnamefont {T.}~\bibnamefont {Driver}}, \bibinfo {author} {\bibfnamefont {N.}~\bibnamefont {Ekanayake}}, \bibinfo {author} {\bibfnamefont {B.}~\bibnamefont {Erk}}, \bibinfo {author} {\bibfnamefont {D.}~\bibnamefont {Garg}}, \bibinfo {author} {\bibfnamefont {E.}~\bibnamefont {Gougoula}}, \bibinfo {author} {\bibfnamefont {D.}~\bibnamefont
  {Heathcote}}, \bibinfo {author} {\bibfnamefont {P.}~\bibnamefont {Hockett}}, \bibinfo {author} {\bibfnamefont {D.~M.~P.}\ \bibnamefont {Holland}}, \bibinfo {author} {\bibfnamefont {A.~J.}\ \bibnamefont {Howard}}, \bibinfo {author} {\bibfnamefont {S.}~\bibnamefont {Kumar}}, \bibinfo {author} {\bibfnamefont {J.~W.~L.}\ \bibnamefont {Lee}}, \bibinfo {author} {\bibfnamefont {S.}~\bibnamefont {Li}}, \bibinfo {author} {\bibfnamefont {J.}~\bibnamefont {McManus}}, \bibinfo {author} {\bibfnamefont {J.}~\bibnamefont {Mikosch}}, \bibinfo {author} {\bibfnamefont {D.}~\bibnamefont {Milesevic}}, \bibinfo {author} {\bibfnamefont {R.~S.}\ \bibnamefont {Minns}}, \bibinfo {author} {\bibfnamefont {S.}~\bibnamefont {Neville}}, \bibinfo {author} {\bibnamefont {{Atia-Tul-Noor}}}, \bibinfo {author} {\bibfnamefont {C.~C.}\ \bibnamefont {Papadopoulou}}, \bibinfo {author} {\bibfnamefont {C.}~\bibnamefont {Passow}}, \bibinfo {author} {\bibfnamefont {W.~O.}\ \bibnamefont {Razmus}}, \bibinfo {author} {\bibfnamefont {A.}~\bibnamefont
  {R{\"o}der}}, \bibinfo {author} {\bibfnamefont {A.}~\bibnamefont {Rouz{\'e}e}}, \bibinfo {author} {\bibfnamefont {A.}~\bibnamefont {Simao}}, \bibinfo {author} {\bibfnamefont {J.}~\bibnamefont {Unwin}}, \bibinfo {author} {\bibfnamefont {C.}~\bibnamefont {Vallance}}, \bibinfo {author} {\bibfnamefont {T.}~\bibnamefont {Walmsley}}, \bibinfo {author} {\bibfnamefont {J.}~\bibnamefont {Wang}}, \bibinfo {author} {\bibfnamefont {D.}~\bibnamefont {Rolles}}, \bibinfo {author} {\bibfnamefont {A.}~\bibnamefont {Stolow}}, \bibinfo {author} {\bibfnamefont {M.~S.}\ \bibnamefont {Schuurman}},\ and\ \bibinfo {author} {\bibfnamefont {R.}~\bibnamefont {Forbes}},\ }\href {https://doi.org/10.1021/acs.jpclett.3c01447} {\bibfield  {journal} {\bibinfo  {journal} {J. Phys. Chem. Lett.}\ }\textbf {\bibinfo {volume} {14}},\ \bibinfo {pages} {7126} (\bibinfo {year} {2023})}\BibitemShut {NoStop}%
\bibitem [{\citenamefont {Barlow}\ \emph {et~al.}(2024)\citenamefont {Barlow}, \citenamefont {Phelps}, \citenamefont {Eng}, \citenamefont {Katayama}, \citenamefont {Sutcliffe}, \citenamefont {Coletta}, \citenamefont {Brechin}, \citenamefont {Penfold},\ and\ \citenamefont {Johansson}}]{barlow_tracking_2024}%
  \BibitemOpen
  \bibfield  {author} {\bibinfo {author} {\bibfnamefont {K.}~\bibnamefont {Barlow}}, \bibinfo {author} {\bibfnamefont {R.}~\bibnamefont {Phelps}}, \bibinfo {author} {\bibfnamefont {J.}~\bibnamefont {Eng}}, \bibinfo {author} {\bibfnamefont {T.}~\bibnamefont {Katayama}}, \bibinfo {author} {\bibfnamefont {E.}~\bibnamefont {Sutcliffe}}, \bibinfo {author} {\bibfnamefont {M.}~\bibnamefont {Coletta}}, \bibinfo {author} {\bibfnamefont {E.~K.}\ \bibnamefont {Brechin}}, \bibinfo {author} {\bibfnamefont {T.~J.}\ \bibnamefont {Penfold}},\ and\ \bibinfo {author} {\bibfnamefont {J.~O.}\ \bibnamefont {Johansson}},\ }\href {https://doi.org/10.1038/s41467-024-48411-0} {\bibfield  {journal} {\bibinfo  {journal} {Nat. Commun.}\ }\textbf {\bibinfo {volume} {15}},\ \bibinfo {pages} {4043} (\bibinfo {year} {2024})}\BibitemShut {NoStop}%
\bibitem [{\citenamefont {Chai}\ and\ \citenamefont {Head-Gordon}(2008)}]{chai_long-range_2008}%
  \BibitemOpen
  \bibfield  {author} {\bibinfo {author} {\bibfnamefont {J.-D.}\ \bibnamefont {Chai}}\ and\ \bibinfo {author} {\bibfnamefont {M.}~\bibnamefont {Head-Gordon}},\ }\href {https://doi.org/10.1039/B810189B} {\bibfield  {journal} {\bibinfo  {journal} {Phys. Chem. Chem. Phys.}\ }\textbf {\bibinfo {volume} {10}},\ \bibinfo {pages} {6615} (\bibinfo {year} {2008})}\BibitemShut {NoStop}%
\bibitem [{\citenamefont {Triguero}\ \emph {et~al.}(1999)\citenamefont {Triguero}, \citenamefont {Plashkevych}, \citenamefont {Pettersson},\ and\ \citenamefont {{\AA}gren}}]{triguero_separate_1999}%
  \BibitemOpen
  \bibfield  {author} {\bibinfo {author} {\bibfnamefont {L.}~\bibnamefont {Triguero}}, \bibinfo {author} {\bibfnamefont {O.}~\bibnamefont {Plashkevych}}, \bibinfo {author} {\bibfnamefont {L.}~\bibnamefont {Pettersson}},\ and\ \bibinfo {author} {\bibfnamefont {H.}~\bibnamefont {{\AA}gren}},\ }\href {https://doi.org/10.1016/S0368-2048(99)00008-0} {\bibfield  {journal} {\bibinfo  {journal} {J. Electron Spectrosc.}\ }\textbf {\bibinfo {volume} {104}},\ \bibinfo {pages} {195} (\bibinfo {year} {1999})}\BibitemShut {NoStop}%
\bibitem [{\citenamefont {Jolly}\ and\ \citenamefont {Hendrickson}(1970)}]{jolly_thermodynamic_1970}%
  \BibitemOpen
  \bibfield  {author} {\bibinfo {author} {\bibfnamefont {W.~L.}\ \bibnamefont {Jolly}}\ and\ \bibinfo {author} {\bibfnamefont {D.~N.}\ \bibnamefont {Hendrickson}},\ }\href {https://doi.org/10.1021/ja00710a012} {\bibfield  {journal} {\bibinfo  {journal} {J. Am. Chem. Soc.}\ }\textbf {\bibinfo {volume} {92}},\ \bibinfo {pages} {1863} (\bibinfo {year} {1970})}\BibitemShut {NoStop}%
\bibitem [{\citenamefont {Zhang}\ \emph {et~al.}(2019)\citenamefont {Zhang}, \citenamefont {Ma}, \citenamefont {Wang}, \citenamefont {Ding}, \citenamefont {Gao}, \citenamefont {Kan},\ and\ \citenamefont {Hua}}]{zhang_accurate_2019}%
  \BibitemOpen
  \bibfield  {author} {\bibinfo {author} {\bibfnamefont {J.-R.}\ \bibnamefont {Zhang}}, \bibinfo {author} {\bibfnamefont {Y.}~\bibnamefont {Ma}}, \bibinfo {author} {\bibfnamefont {S.-Y.}\ \bibnamefont {Wang}}, \bibinfo {author} {\bibfnamefont {J.}~\bibnamefont {Ding}}, \bibinfo {author} {\bibfnamefont {B.}~\bibnamefont {Gao}}, \bibinfo {author} {\bibfnamefont {E.}~\bibnamefont {Kan}},\ and\ \bibinfo {author} {\bibfnamefont {W.}~\bibnamefont {Hua}},\ }\href {https://doi.org/10.1039/c9cp04573b} {\bibfield  {journal} {\bibinfo  {journal} {Phys. Chem. Chem. Phys.}\ }\textbf {\bibinfo {volume} {21}},\ \bibinfo {pages} {22819} (\bibinfo {year} {2019})}\BibitemShut {NoStop}%
\bibitem [{\citenamefont {Ge}\ \emph {et~al.}(2022)\citenamefont {Ge}, \citenamefont {Zhang}, \citenamefont {Wang}, \citenamefont {Wei},\ and\ \citenamefont {Hua}}]{ge_qmmm_2022}%
  \BibitemOpen
  \bibfield  {author} {\bibinfo {author} {\bibfnamefont {G.}~\bibnamefont {Ge}}, \bibinfo {author} {\bibfnamefont {J.-R.}\ \bibnamefont {Zhang}}, \bibinfo {author} {\bibfnamefont {S.-Y.}\ \bibnamefont {Wang}}, \bibinfo {author} {\bibfnamefont {M.}~\bibnamefont {Wei}},\ and\ \bibinfo {author} {\bibfnamefont {W.}~\bibnamefont {Hua}},\ }\href {https://doi.org/10.1021/acs.jpcc.2c05405} {\bibfield  {journal} {\bibinfo  {journal} {J. Phys. Chem. C}\ }\textbf {\bibinfo {volume} {126}},\ \bibinfo {pages} {15849} (\bibinfo {year} {2022})}\BibitemShut {NoStop}%
\bibitem [{siC()}]{siC18}%
  \BibitemOpen
  \href@noop {} {}\bibinfo {note} {See Supplemental Material at [URL will be inserted by publisher] for computational details; PESs by M06-2X; HOMO-LUMO gap and the lowest C1s orbital energy by \ensuremath{\omega}B97XD; Supplementary data of NEXAFS spectra.}\BibitemShut {Stop}%
\bibitem [{\citenamefont {Hua}\ \emph {et~al.}(2010)\citenamefont {Hua}, \citenamefont {Gao}, \citenamefont {Li}, \citenamefont {{\r A}gren},\ and\ \citenamefont {Luo}}]{hua_x-ray_2010}%
  \BibitemOpen
  \bibfield  {author} {\bibinfo {author} {\bibfnamefont {W.}~\bibnamefont {Hua}}, \bibinfo {author} {\bibfnamefont {B.}~\bibnamefont {Gao}}, \bibinfo {author} {\bibfnamefont {S.}~\bibnamefont {Li}}, \bibinfo {author} {\bibfnamefont {H.}~\bibnamefont {{\r A}gren}},\ and\ \bibinfo {author} {\bibfnamefont {Y.}~\bibnamefont {Luo}},\ }\href@noop {} {\bibfield  {journal} {\bibinfo  {journal} {Phys. Rev. B}\ }\textbf {\bibinfo {volume} {82}},\ \bibinfo {pages} {155433} (\bibinfo {year} {2010})}\BibitemShut {NoStop}%
\bibitem [{\citenamefont {Ge}\ \emph {et~al.}(2024)\citenamefont {Ge}, \citenamefont {Zhang}, \citenamefont {Wang}, \citenamefont {Wei}, \citenamefont {Ji}, \citenamefont {Duan}, \citenamefont {Ueda},\ and\ \citenamefont {Hua}}]{ge_mapping_2024}%
  \BibitemOpen
  \bibfield  {author} {\bibinfo {author} {\bibfnamefont {G.}~\bibnamefont {Ge}}, \bibinfo {author} {\bibfnamefont {J.-R.}\ \bibnamefont {Zhang}}, \bibinfo {author} {\bibfnamefont {S.-Y.}\ \bibnamefont {Wang}}, \bibinfo {author} {\bibfnamefont {M.}~\bibnamefont {Wei}}, \bibinfo {author} {\bibfnamefont {Y.}~\bibnamefont {Ji}}, \bibinfo {author} {\bibfnamefont {S.}~\bibnamefont {Duan}}, \bibinfo {author} {\bibfnamefont {K.}~\bibnamefont {Ueda}},\ and\ \bibinfo {author} {\bibfnamefont {W.}~\bibnamefont {Hua}},\ }\href {https://doi.org/10.1021/acs.jpclett.4c01133} {\bibfield  {journal} {\bibinfo  {journal} {J. Phys. Chem. Lett.}\ }\textbf {\bibinfo {volume} {15}},\ \bibinfo {pages} {6051} (\bibinfo {year} {2024})}\BibitemShut {NoStop}%
\bibitem [{\citenamefont {Wei}\ \emph {et~al.}(2026)\citenamefont {Wei}, \citenamefont {Liu}, \citenamefont {Wang}, \citenamefont {Zhang}, \citenamefont {Zhang}, \citenamefont {Ge},\ and\ \citenamefont {Hua}}]{wei_c18_data_2026}%
  \BibitemOpen
  \bibfield  {author} {\bibinfo {author} {\bibfnamefont {M.}~\bibnamefont {Wei}}, \bibinfo {author} {\bibfnamefont {Z.}~\bibnamefont {Liu}}, \bibinfo {author} {\bibfnamefont {S.-Y.}\ \bibnamefont {Wang}}, \bibinfo {author} {\bibfnamefont {J.-R.}\ \bibnamefont {Zhang}}, \bibinfo {author} {\bibfnamefont {L.}~\bibnamefont {Zhang}}, \bibinfo {author} {\bibfnamefont {G.}~\bibnamefont {Ge}},\ and\ \bibinfo {author} {\bibfnamefont {W.}~\bibnamefont {Hua}},\ }\href@noop {} {\bibinfo {title} {Data for ``{Core-Level Spectroscopy Decodes Bond-Alternation Dynamics of Cyclo[18]Carbon}''}} (\bibinfo {year} {2026}),\ \bibinfo {note} {version 1.0, [Dataset], Zenodo, doi:10.5281/zenodo.21851536}\BibitemShut {NoStop}%
\end{thebibliography}
%\end{CJK*}

\end{document}